\documentclass[12pt]{article}

\newif\ifpdf
\ifx\pdfoutput\undefined
\pdffalse % we are not running PDFLaTeX
\else
\pdfoutput=1 % we are running PDFLaTeX
\pdftrue
\fi

\ifpdf
\usepackage[pdftex]{graphicx}
\else
\usepackage{graphicx}
\fi

\newcommand {\vp} {\varphi}
  \newcommand {\be} {\begin{equation}}
  \newcommand {\ee} {\end{equation}}
  \newcommand {\bea} {\begin{eqnarray}}
  \newcommand {\eea} {\end{eqnarray}}

\begin{document}

\title{\noindent  Some General Results for Multi-dimensional Compactons
in Generalized $N$-dimensional KdV Equations}
\author{Fred Cooper$^{\ast\dag}$, Avinash Khare$^{\ast\ast}$, and Avadh 
Saxena$^{\dag}$}
%\email{cooper@santafe.edu}

\maketitle

%\author{Avinash Khare}
%\email{khare@iopb.res.in}

{$^\ast$Santa Fe Institute, Santa Fe, NM 87501, USA}

{$^\dag$Theoretical Division and Center for Nonlinear Studies, Los Alamos
National Laboratory, Los Alamos, NM 87545, USA}

{$^{\ast\ast}$Institute of Physics, Bhubaneswar, Orissa 751005, India}

\begin{abstract}
We derive a general theorem relating the energy and momentum
with the velocity of any solitary wave solution of the generalized
KdV equation in $N$-dimensions that follows from an action principle. 
Further, we show that our $N$-dimensional Lagrangian formulation leads 
to a subclass of the equations discussed recently by Rosenau, Hyman 
and Staley.
\end{abstract}

\newpage

\section{Introduction} 
Recently, in an interesting paper, Rosenau, Hyman and Staley \cite {RHS} 
have shown the existence of spherically symmetric compactons in arbitrary
number of dimensions but traveling uniformly in $x$ direction. The purpose
of this article is to show that a subclass of these compacton equations 
(in arbitrary number of dimensions) can be derived from a Lagrangian. For
this subclass, we derive a general theorem relating the energy and momentum, 
with the velocity of any solitary wave solution even if the solution is 
not known in a closed form, and using this information comment about the 
stability of this subclass of compacton solutions.

Compactons were originally discovered by Rosenau and Hyman (RH) \cite{RH} 
in their investigation of a class of generalized KdV equations described 
by the parameters $(m,n)$.
  \be\label{1}
   u_t + (u^m)_x + (u^n)_{xxx} = 0 . 
  \ee
  They found \cite{RH,H} among other things, that these equations 
  have  solitary wave solutions of  the form
  $A [\cos(d\xi)]^{2/(m-1)}$ for $m=2,3$,
  where $\xi= x-ct$ and $ -\pi/2  \leq d  \xi \leq \pi/2$.
  However these systems of 
  equations were not in general derivable from a 
  Lagrangian. For that reason, Cooper, Shepard and Sodano (CSS)  \cite{bib:kdv1} were  led to consider a related system of equations derivable from
  a Lagrangian. 
  By starting with the  first order Lagrangian 
  \cite{bib:kdv1,bib:khare} 
  \be\label{2}
  {\cal L}(l,p) = \int \left(  \frac{1}{2} \vp_{x} \vp_{t} + {
  {(\vp_{x})^{l}} \over {l(l-1)}}  -\alpha (\vp_{x})^{p} (\vp_{xx})^{2}  \right)
  dx \equiv \int L dx\,, 
  \ee
CSS  derived and studied  a generalized sequence of KdV equations
  of the form:
  \be\label{3}
   u_t + u^{l-2} u_x + \alpha [2u^p  u_{xxx} + 4p
  u^{p-1} u_x u_{xx} + p(p-1) u^{p-2} (u_x)^3 ] =0\,,
  \ee 
  where the  usual field $u(x,t)$
  of the generalized KdV equation is defined by $u(x,t) = \vp_{x}(x,t)$.
  The equations of CSS had the same class of solutions as the original 
  RH equations but had the advantage of being derived
  from an action and thus conserved energy.  The CSS equations lead to similar
  compacton solutions when the parameters $(m,n)$ of the RH equation have the
  value $(l-1,p+1)$.  In general the RH and CSS equations have {\em different} conservation laws.
  
  Recently, Rosenau, Hyman and Staley \cite{RHS} have shown that
  a generalization of their original set of equations, namely
  \be\label{4}
 { \cal{C_N}}(m,a+b): u_t+(u^m)_x + \frac{1}{b} [u^a(\nabla^2 u^b)]_x
 =0\,,
 \ee 
  admitted traveling compacton solutions which were spherically symmetric but 
  traveling uniformly in $x$ direction.  Here $m$, $a$ and $b$ are integers. 
Introducing the variables:
  \be\label{5}
  s=x-\lambda t; ~~ R= \sqrt{s^2+x_{2}^2+...+x_{N}^2}\,,
  \ee
  one can integrate their equation in a traveling frame to find 
  \be\label{6}
  u^a \left[-\lambda u^{1-a}+ u^{m-a} +
  \frac{1}{b R^{N-1} } \frac{d}{dR} R^{N-1} \frac{d}{dR} u^b\right] =0\,.
  \ee
Let us assume that the general solution of this equation is of the form
\be\label{7}
u= A U[\beta R]\,,
\ee
and further let the velocity dependence of $A$ and $\beta$ be  
\be\label{7a}
A \propto \lambda^{\sigma}\,,~~\beta \propto \lambda^{\eta}\,.
\ee
Then one can immediately scale out $\lambda$ provided
\be\label{8}
\sigma=\frac{1}{m-1}; ~~ 
\eta = \frac{m-n}{2(m-1)}\,,
\ee
where $n=a+b$. It is amusing to note that the velocity dependence of both
the amplitude $A$ and the width $\beta$ is dimension independent.
Further, when $m=n$ the width is independent of the velocity of the compacton.
(Note that $N$ denotes the number of space dimensions which is
different from $n=a+b$). From the RHS Eq. (\ref{4}) it immediately follows
that one of the conserved quantity in the RHS model is the mass $M$
defined by
\be\label{7b}
M= \int u d^{N}x\,.
\ee
Using Eqs. (\ref{7a}) and (\ref{8}) it immediately follows that the
velocity dependence of $M$ is however dimension dependent
\be\label{7c}
M \propto \lambda^{[N(n-m)+2]/2(m-1)}\,.
\ee
Nevertheless, for compactons since $m=n$, this $N$-dependence disappears 
even for $M$.

RHS found two explicit solutions in $N$-dimensions: (a) true compacton 
solution in the case $a=1$, i.e. ${ \cal{C_N}}(m=1+b$, $a+b=1+b)$ and  
(b) ${ \cal{C_N}}(m=2,$ $n=a+b=3)$ when the solution is part of a parabola.

We shall now show that a subclass of RHS Eq. (\ref{4}) satisfying
$b=a+1$ can be derived from a generalization of the CSS Lagrangian (\ref{2}) 
with $(\vp_{xx})^2 \rightarrow (\nabla\phi_x\cdot\nabla\phi_x)$.
In particular, consider the Action  
  \be\label{9} 
  S=\int L(l,p) \,dx dt = \int \left(  \frac{1}{2} \vp_{x} \vp_{t} + {
  {(\vp_{x})^{l}} \over {l(l-1)}} - \alpha (\vp_{x})^{p} 
  ( \nabla \vp_{x} \cdot \nabla \vp_x )  \right)\,dx dt \,.
  \ee
~ From this we obtain the equation:
\be\label{10}
u_t +(\partial _x)[u^{l-1}/(l-1)]+\alpha (\partial_x) [p  u^{p-1}  \nabla u \cdot \nabla u]
+\alpha (\partial_ x) [ 2 u^p \nabla^2 u]=0\,.
\ee
Although this is the natural generalization of the CSS equation, at this 
point to make the comparison with RHS more explicit, we will modify the 
coefficients of the Lagrangian density to be:
  \be\label{11} 
  L_2(l,p) = \frac{1}{2} \vp_{x} \vp_{t} + \frac {(\vp_{x})^l} {l} - 
\alpha (\vp_{x})^{p} ( \nabla \vp_{x} \cdot \nabla \vp_x)  \,.
  \ee
  The equation of motion that results from this can be written in the form:
  \be\label{12}
u_t +(\partial_x)[u^{l-1}] 
+\frac{4 \alpha}{(p+2)} \partial_x  [ u^{p/2}  \nabla^2 u^{(p/2+1)}]=0\,.
\ee
Thus by identifying 
\be\label{13}
a=p/2\,, ~~ l=m+1\,,	~~ b=a+1\,,~~ \alpha=1/2\,,
\ee
we obtain a subclass of the RHS equation with $b=a+1$.  We have not found 
a way to generalize the CSS Lagrangian to obtain the full family of RHS 
equations with arbitrary $b$ which is independent of $a$.  For $b=a+1$, 
the Lagrangian (\ref{11}) was also recently written down in the Appendix 
of \cite{PR}.  

\section{Conservation Laws}
~From Eq. (\ref{12}) we find that
\be\label{15}
M = \int u~ d^{N}x\,, 
\ee
is conserved.
This conservation law also follows from the invariance of the action under 
$\vp \rightarrow \vp+\vp_0$, where $\vp_0$ is a constant. 

On multiplying Eq. (\ref{12}) by $u$ it is easily shown that another
conserved quantity is $P= \frac{1}{2} \int u^2~d^{N}x$. In particular, the
corresponding continuity equation is of the form
\be\label{16}
\frac{\partial \rho}{\partial t}+\frac{\partial j_k}{\partial x_k}=0\,.
\ee
Here summation over $k$ is understood with $k$ counting the number of
space indices. In this case the relevant matter density $\rho$ and
the current density $j_k$ are given by
\bea\label{17}
&&\rho =\frac{u^2}{2}\,,~j_y=\frac{2\alpha}{p+1}u^{p+1}\frac{\partial^2 u}
{\partial x \partial y}\,,
~~j_z=\frac{2\alpha}{p+1}u^{p+1}\frac{\partial^2 u}
{\partial x \partial z}\,, ... \nonumber \\
&&j_x=\frac{(l-1)u^l}{l}+\frac{2\alpha p}{(p+1)} u^{(p+1)} \nabla^2 u
+\frac{2\alpha}{(p+1)} u^{(p+1)} \frac{\partial^2 u}{\partial x^2}
+\alpha (p-1)u^{p}(\nabla u \cdot \nabla u)\,.
\eea
A third conserved quantity is obviously the Hamiltonian corresponding to 
the Lagrangian density $L_2$ as given by Eq. (\ref{11}). It is given by
\be\label{18}
H = \int d^N x \left[ -  \frac{(\vp_{x})^l} {l} + \alpha (\vp_{x})^{p} 
( \nabla \vp_{x} \cdot \nabla \vp_x) \right]\,.
\ee
More generically, the Lagrangian density $L_2$ in Eq. (\ref{11}) is of 
the form:
\be
L= L( \phi_\mu, \phi_{\mu \nu}). 
\ee
Here we will use the compact notation that $\mu$ runs from $0, 1 \ldots N$ 
with $x_0=t$ and $x_1=x$.  Also $\phi_\mu \equiv \partial \phi/\partial 
x_\mu$, etc. The stationary condition on the action under variations of 
$\phi$ of the generic form $\phi \rightarrow \phi+ \delta \phi$, with $\delta 
\phi$ vanishing on the boundaries of the integration range, leads to the 
Euler-Lagrange equation for this type of Lagrangian:
\be
\partial_\mu \frac {\delta L}{\delta \phi_\mu} = \partial_ \mu \partial_ 
\nu \frac {\delta L}{\delta \phi_{\mu \nu}} . 
\ee
Here we use the Einstein summation convention that repeated indices are summed over.
To obtain the energy-momentum tensor we consider an infinitesimal arbitrary coordinate transformation
\be
x_\mu \rightarrow x_\mu + \epsilon_\mu (x) . 
\ee
Under this transformation 
\be 
\phi(x) \rightarrow \phi(x+\epsilon(x)) \equiv \phi(x) + \delta \phi(x) , 
\ee
which for infinitesimal transformations leads to 
\be
\delta \phi(x) = \phi_\mu  \epsilon_\mu . 
\ee

Under this transformation we obtain:
\be
\delta S = \int d^{N+1} x \left[ \frac  {\delta L} {\delta \phi_\mu} \partial_\mu  ( \phi_\lambda  \epsilon_\lambda ) +
\frac  {\delta L} {\delta \phi_{\mu \nu} } \partial_\mu \partial_\nu ( \phi_\lambda  \epsilon_\lambda)  \right].
\ee
One identifies the energy-momentum tensor $T_{\mu \lambda}$ from the coefficient 
of $\partial_\mu \epsilon_\lambda$. That is after integration by parts and using 
the Euler-Lagrange equation one can cast $\delta S$ in the form
\be
\delta S = \int d^{N+1} x \left[ T_{\mu \lambda} \partial_\mu \epsilon_\lambda 
\right] ,  
\ee
where
\be
 T_{\mu \lambda} = -\delta_{\mu \lambda} L + \frac{\delta L} {\delta \phi_\mu}  \phi_\lambda
 + 2 \frac{\delta L} {\delta \phi_{\mu \nu}} \phi_{\nu \lambda} - \partial_\nu \left( \frac{\delta L} {\delta \phi_{\mu \nu}} \phi_\lambda \right).
 \ee
 The $N+1$ conservation laws that arise from the fact that
 \be
 \partial_\mu T_{\mu \lambda} = 0
 \ee
 are  $dP_{\mu} /dt =0$ where
 \be
 P_\mu = \int d^N x T_{0 \mu}.
 \ee
 The conserved energy $H$ is $P_0$ and there are N conserved linear momenta $P_k$. 
 In particular for the Lagrangian considered in this paper, the only fields entering the Lagrangian are
 $\phi_0$, $\phi_x$, and $\phi_{x k}$.
 This results in the fact that
 \be T_{00} = -L + \frac {1}{2} \phi_x \phi_t = {\cal H} \ee
 and
 \be
 T_{0k} =  \frac {1}{2} \phi_x \phi_k; ~~k= 1~ \ldots ~N.
 \ee
Note that $\vp_k$, $k\ne1$ is only known through an integral operator, see 
Eq. (36) below.  The tensor $T_{ik}$ can be constructed by using 
\be \delta L/\delta\vp_i 
= \delta_{i1}[\vp_t/2+(\vp_x)^{l-1}/(l-1)-\alpha p (\vp_x)^{p-1}\nabla\vp_x 
\cdot\nabla\vp_c] 
\ee 
and 
\be \delta L/\delta\vp_{i\nu}=-2\alpha\delta_{i1} 
\delta_{\nu k}\vp_{xk}(\vp_x)^p. 
\ee 
Because the compactons have a preferred $x$ direction, there is also a 
rotational symmetry about the $x$ axis leading to conservation of angular 
momentum about that axis.  

We shall see below that in general there are only two conserved quantities
corresponding to the RHS Eq. (\ref{4}).

\section{General Properties of the Exact Solutions}

Let us assume that the general solution to the field Eq. (\ref{12})
is of the form
\be\label{19}
u \equiv \vp_x = A Z (\beta R)\,,
\ee
where
$A$ and $\beta$ are constants, $Z$ is an unspecified function and 
\be\label{20}
R= \left[ ((x+q(t))^2 + \sum_{i=2}^N x_k^2 \right]^{1/2}\,.
\ee
Using the fact that $\beta$ must be chosen to minimize the action we
shall deduce the general behavior of the conserved quantities $M$, $P$ 
and $H$ as well as the amplitude $A$ and the width $\beta$ as a function 
of the velocity ${\dot q}$. 

To evaluate the action we must first determine $\vp$.  A convenient choice  for the lower
value limit of integration (see also \cite{kdv1}) leads to:
\be\label{21}
\vp(x, x_i, t)  = \int^x_{-q(t)}  dx' \left[ A Z ( \beta R(x'+q(t), x_i)
\right]\,.
\ee
Introducing the rescaled variables $y'=\beta(x'+q(t))$, $y_i= \beta x_i$,
 we have
\be
\vp(x,t)= \frac{A}{\beta} \int_0^{\beta(x+q(t))} ~dy~ Z[( y^2+\sum y_i^2)^{1/2}]. 
\ee
Thus we find that
\be\label{22}
\vp_t = {\dot q} ~\vp_x\,,
\ee
and the the kinetic term in the action is 
\be\label{23}
KE = \int d^Nx \frac{1}{2}  \vp_x \vp_t 
=  {\dot q} \int d^Nx \frac{1}{2} ( \vp_x )^2 = {\dot q} P\,.
\ee
Thus the  effective Lagrangian has the canonical form for point mechanics:
\be\label{24}
L = P {\dot q} - H\,, 
\ee
 where 
\be\label{25}
H = \int d^Nx\, {\cal H}\,,
\ee
is evaluated for the field ansatz.

We can now eliminate $A$ in all the conservation equations 
in favor of the conserved momentum $P$.
Defining the constants $C_l$ via
\be\label{26}
C_l =\int dr ~~ Z^l (r) r^{N-1} \Omega_N\,, 
\ee
where $\Omega_N$ is the solid angle in $N$ dimensions, we find that
\be\label{27}
P = \frac {A^2}{2 \beta^N} C_2\,,
\ee
so that
\be\label{28}
A = \sqrt{\frac{2}{C_2}} \beta^{N/2} P^{1/2}\,.
\ee
The constant of motion $M$ is given by 
\be\label{29}
M= \int d^N x ~u = \sqrt{\frac{2}{C_2}} ~C_1~\frac { P^{1/2} }{ \beta^{N/2}}\,.
\ee

The key point to note is that 
$\beta$  can be determined in terms of $P$ 
using the fact that the actual solution of Hamilton's equations satisfies 
\be\label{30}
\frac {\partial H} {\partial \beta} = 0\,. 
\ee
This is a consequence of the fact that if we consider a reduced class of trial 
functions parametrized by $\beta$ then the solution that minimizes the energy  
satifies Eq. (\ref{30}).  When the actual solution is in the class of trial 
functions, then the value of $\beta$ that leads to a solution of the actual 
equations of motion is the value of $\beta$ obtained from the minimization of 
the energy.   This is due to the fact that the exact equations of motion 
minimize the Action.

For the Hamiltonian we find:
\be\label{31}
H= -f_1(l) P_x^{l/2} \beta^{N(l-2)/2} + f_2(p) P^{(p+2)/2}
\beta^{(Np+4)/2}\,,
\ee
where 
\be\label{32}
f_1(l) = \frac{C_l}{l}\left( \frac{2}{C_2} \right)^{l/2}\,, 
~~~~f_2(p) = \alpha D_p \left( \frac{2}{C_2} \right)^{(p+2)/2}\,,
\ee
and
\be\label{33}
D_p = \int dr ~ r^{N-1}[Z(r)]^p \left(dZ/dr \right)^2 \Omega_N\,.
\ee
Minimizing $H$ with respect to $\beta$ we obtain:
\be\label{34}
\beta = K(N,p,l)^{2/[N(l-p-2)-4]} P^{(p-l+2)/[N(l-p-2)-4]}\,,
\ee
where
\be\label{35}
K(N,p,l) = \frac{f_2(p) (Np+4)}{N(l-2) f_1(l)}\,.
\ee
Therefore  
\be\label{36}
H= \frac{N(l-p-2)-4}{Np+4} f_1(l) K^{N(l-2)/[N(l-p-2)-4]} P^r\,,
\ee
where
\be\label{37}
r= \frac{(p+2)N -l(N-2)}{N(p+2-l)+4}\,.
\ee
Note that for $N=1$ we get our previous result that
\be\label{38}
r = \frac{p+2+l}{6+p-l}\,.
\ee
Using
\be\label{39}
\lambda \equiv {\dot q} = \frac {\partial H}{\partial P}\,,
\ee
then we find, in analogy with the CSS equation, that we have 
\be\label{40}
\lambda = r  \frac {H}{P}\,,
\ee
where $r$ is now dependent on $N$ and given by Eq. (\ref{38}).
We also find that 
\be\label{41}
P \propto \lambda^{~[N(p+2-l)+4]/ [2(l-2)]}
\ee
and
\be\label{42}
\beta \propto  \lambda^{~(l-p-2)/ [2(l-2)]}\,,
~~A \propto  \lambda^{~1/ [(l-2)]}\,.
\ee
In addition, we find that
\be\label{43}
M \propto \lambda^{~[N(p+2-l)+2]/ [2(l-2)]}\,. 
\ee
It is worth noting that while the dependence of  $M$ and $P$ on $\lambda$ 
(i.e., ${\dot q}$) is dimension-dependent, this is not so for $A$ and 
$\beta$. Note however that in the special case of compactons since $l=p+2$ 
hence, in that case, the dependence of $H$, $P$ and $M$ on $\lambda$ is 
also independent of the number of dimensions $N$. On using the 
correspondence $l=m+1$, $p=n-1=a+b-1$ between RHS and our model, it is 
easy to check that, as expected, the $\lambda$ dependence of $A$, $\beta$ 
and $M$ is exactly as given by Eqs. (\ref{8}) and (\ref{7c}).
Note however that without knowing the Lagrangian, one cannot say anything
about the velocity dependence of energy $H$ and momentum $P$ which is
required to study the question about the stability of such solutions.

{\bf Stability of Solutions}: Following the analysis in \cite{dk}, it is
clear that the criterion for linear stability is equivalent to the 
condition
\be
\frac{\partial P}{\partial \lambda} > 0\,.
\ee
Using Eq. (\ref{41}), it then follows that the compacton solutions of our
model (i.e. RHS model with $b=a+1$) in $N$-dimensions are stable provided  
$2 < l < p+2+4/N$. However, it is not possible to say anything about the 
other compacton solutions of the RHS model.

\section{Comparison and Contrast Between General RHS \\ Equation 
and the Lagrangian Subset}

Before finishing this article, we would like to make several remarks
about the RHS model \cite{RHS} {\it vis a vis} the subset of models 
defined by our Lagrangian. 
\begin{enumerate}

\item While there are $N+2$ conserved quantities (excluding angular momentum 
about the $x$ axis) in our model with $b=a+1$, (i.e. H, $\int u d^{N}x$ and 
$\int u^2 d^{N} x$), there are in general only two known conserved quantities 
(i.e. $\int u d^{N} x$ and $\int u^{b-a+1} d^{N} x$) in the RHS model. While 
the conservation of $\int u d^{N}x$ is rather obvious from the RHS field 
Eq. (\ref{4}), the conservation of $\int u^{b-a+1} d^Nx$ is not so obvious. 
However, one can show that the continuity equation (\ref{16}) is satisfied 
in that case with
\bea\label{45}
&&\rho =\frac{u^{b+1-a}}{b+1-a}\,,~j_y=\frac{b-a}{2b-1}u^{2b-1}
\frac{\partial^2 u}{\partial x \partial y}\,,
~~j_z=\frac{b-a}{2b-1}u^{2b-1}\frac{\partial^2 u}
{\partial x \partial z}\,, ... \nonumber \\
&&j_x=\frac{mu^{(m+b-a)}}{(m+b-a)} +\left(1-\frac{a}{b}\right) 
\bigg [\frac{b}{(2b-1)} u^{(2b-1)} \frac{\partial^2
u}{\partial x^2} \nonumber \\
&&+\frac{(a+b-2)}{2}u^{(2b-2)}(\nabla u \cdot \nabla u)
+\frac{(a+b-1)}{(2b-1)} u^{(2b-1)} \nabla^2 u \bigg ]\,.
\eea
In the special case when $b=a+1$, then as expected this expression
reduces to the one given by Eq. (\ref{17}) provided we make use of the
identification as given by Eq. (\ref{13}).

It may be noted that in case $b-a+1=0$ then one can show in the RHS model 
that a third conserved quantity is $\ln u$. In particular, in that case 
the continuity Eq. (\ref{16}) is satisfied with
\bea\label{46}
&&\rho =\ln u\,,~j_y=u^{(2b-2)}\frac{\partial u}{\partial x}\frac{\partial
u}{\partial y}\,,
~j_z=u^{(2b-2)}\frac{\partial u}{\partial x}\frac{\partial
u}{\partial z}\,,... \nonumber \\
&&j_x=\frac{mu^{(m-1)}}{(m-1)}+u^{(2b-1)} \nabla^2 u
+u^{(2b-2)} \left(\frac{\partial u}{\partial x}\right)^2 
+\frac{(2b-3)}{2}u^{(2b-2)}(\nabla u \cdot \nabla u)\,.  
\eea
 
Similarly, when $b+1-a=1$, one can show that a third conserved quantity
is $u\ln u$. In particular, continuity Eq. (\ref{16}) is satisfied with
\bea\label{47}
&&\rho =u\ln u\,,~j_y=-u^{(2b-2)}\frac{\partial u}{\partial x}\frac{\partial
u}{\partial y}\,,
~j_z=-u^{(2b-2)}\frac{\partial u}{\partial x}\frac{\partial
u}{\partial z}\,, ...\nonumber \\
&&j_x=u^{m}\ln u+\frac{(m-1)u^{m}}{m}+(\ln u+1)
u^{(2b-1)} \nabla^2 u \nonumber \\
&&-u^{(2b-2)} \left(\frac{\partial u}{\partial x}\right)^2 +\left[\frac{ 
(2b-1)}{2}+(b-1)\ln u\right]u^{(2b-2)}(\nabla u \cdot \nabla u)\,.
\eea

\item In the special case when $a=0$ and $b=m$, one can easily show that
the RHS model then has two more conserved quantities given by $\int u\cos 
(\sqrt{b}x)d^Nx$ and $\int u\sin(\sqrt{b}x)d^Nx$. In particular, the 
corresponding densities are \bea\label{48}
&&\rho =u\cos(\sqrt{b}u)\,,~j_y=bu^{(b-1)}\frac{\partial u}{\partial y}
\sin(\sqrt{b}x)\,,
~j_z=bu^{(b-1)}\frac{\partial u}{\partial z}\sin(\sqrt{b}x)\,,... \nonumber \\
&&j_x=-u^{b}\cos(\sqrt{b}x)+\frac{1}{b}\cos(\sqrt{b}x) \nabla^2 u^{b}
+\sqrt{b}u^{(b-1)} \frac{\partial u}{\partial x}\sin(\sqrt{b}x)\,, 
\eea
\bea\label{49}
&&\rho =u\sin(\sqrt{b}u)\,,~j_y=bu^{(b-1)}\frac{\partial u}{\partial y}
\cos(\sqrt{b}x)\,,
~j_z=bu^{(b-1)}\frac{\partial u}{\partial z}\cos(\sqrt{b}x)\,,... \nonumber \\
&&j_x=-u^{b}\sin(\sqrt{b}x)+\frac{1}{b}\sin(\sqrt{b}x) \nabla^2 u^{b}
-\sqrt{b}u^{(b-1)} \frac{\partial u}{\partial x}\cos(\sqrt{b}x)\,. 
\eea
We believe that the existence of these two extra conserved
quantities could perhaps explain the numerical observation of RHS 
as to why $a=0$, $m=n$ compactons are much closer to being elastic 
compared to those compactons with $m=n$ but $a \ne 0$.

\item In the case of the RHS model, the compacton solutions are easily
written down in $N$-dimensions in case $a=1$, $m=n=b+1$. In particular, 
following RHS \cite{RHS}, if we write down the solution as 
\be\label{50}
u^{b} = \lambda \left[1-\frac{F(R)}{F(R_{*})}\right]\,,~~ 0 < R < R_{*}\,,
\ee
and vanishing elsewhere, then one can show that the solution in
$N$-dimensions is 
\bea\label{51}
&&F(R) = R^{-k}J_{k}(\sqrt{b}R)\,,~ k=(N-2)/2=0,1,2,...\,, \nonumber \\
&&F(R) = \left[\frac{1}{R}\frac{d}{dR}\right]^{k} [c_1
\sin(\sqrt{b}R)+c_2\cos(\sqrt{b}R]\,,~~k=(N-3)/2=0,1,2,...\,
\eea
where $J_{n}(x)$ is the Bessel function of order $n$ \cite{GR}. 

\item Similarly, it is easy to generalize the parabolic compacton solution 
of RHS in the case $m=2$, $n=a+b=3$.  In particular, it is easily shown 
that in general, if $m=(n+1)/2$ then the parabolic compacton solution to 
Eq. (\ref{4}) is given by
\be\label{52}
u^{(n-1)/2}= A-BR^2\,,~~0 < R \le R_{*} \equiv \sqrt{A/B} , 
\ee
where
\be\label{53}
A=\frac{\lambda [N(m-1)+b+1-a]}{b+1-a}\,,~~
B=\frac{(m-1)^2}{2b[N(m-1)+b+1-a]}\,.
\ee

\item Apart from these two compacton solutions, it is also possible to
obtain other compacton solutions to RHS Eq. (\ref{4}).
For example, on using the ansatz
\be\label{54}
s=x-\lambda t+\alpha_2 y+\alpha_3 z+...\,,
\ee
in Eq. (\ref{4}), it is easily shown that the RHS Eq. (\ref{4})
essentially reduces to a one-dimensional problem. In particular, on
integrating twice, 
and assuming the two constants of integration to be
zero, one can show that the RHS Eq. (\ref{4}) takes the form 
\be\label{55}
\left(\frac{du}{ds}\right)^2 =\frac{2\lambda u^{3-n}}{b(b+1-a)C}
-\frac{2u^{m+2-n}}{b(m+b-a)C}\,,
\ee
where $C=1+\alpha_2^2+\alpha_3^2+...$. This is a one-dimensional equation
which represents the generalization of the old RH equation \cite{RH}.
Following our earlier work \cite{CKS}, one can immediately write down 
the compacton as well as the elliptic and even general hyper-elliptic 
compacton solutions of Eq. (\ref{55}). For example, for $m=n$, the 
compacton solution is given by
\be\label{56}
u=A[\cos(\beta s)]^{2/(n-1)}\,,~~-\pi /2 \le \beta s \le \pi /2\,,
\ee
and zero elsewhere, where
\be\label{57}
A=\left[\frac{2b\lambda}{b+1-a}\right]^{1/(n-1)}\,,~\beta=\frac{(n-1)} 
{2b\sqrt{C}}
\ee
On the other hand, for $m=2n-1$, the elliptic compacton solution is given
by
\be\label{58}
u=A \big [cn(\beta s,k^2=1/2) \big ]^{2/(n-1)}\,,
~~-K(k^2=1/2) \le \beta s \le K(k^2=1/2)\,,
\ee
and zero elsewhere, where
\be\label{59}
A=\left[\frac{(3b+a-1)\lambda}{b+1-a} \right]^{1/2(n-1)}\,,
~\beta=\frac{(n-1)}{\sqrt{bC}}\left[\frac{\lambda}{(b+1-a)(3b+a-1)} 
\right]^{1/4}\,.  
\ee
Here $cn(x,k)$ denotes a Jacobi elliptic function with modulus $k$ and 
$K(k)$ is the complete elliptic integral of the first kind \cite{GR}.  
Similarly, the other solution of \cite{CKS} (when $m=3n-2$) can be written 
down as well. 

\end{enumerate}

\newpage 

\section{Conclusion} 
In this article we have generalized the Lagrangian of CSS \cite{bib:kdv1} 
to $N$ dimensions and have obtained a system of KdV equations which are a 
subset of the recently discovered equations of RHS \cite{RHS} which support 
$N$-dimensional compactons. By exploiting the Lagrangian structure we were 
able to obtain certain general relations among the conserved quantities 
(derived from an energy-momentum tensor) and how these quantities depend 
on the velocity of a traveling wave solution. Using this information we 
remarked on the question of the stability of such solutions, specifically 
one would expect the $a=0$ compactons to be more stable due to the existence 
of two more ($N+2$) conserved quantities. 

This work was supported in part by the U.S. Department of Energy.  We would 
like to thank J. M. Hyman for sharing his research with us prior to its 
publication and for fruitful discussions.

%\newpage

 \end{document}